\documentclass[twocolumn,pra,aps,superscriptaddress,showpacs, tightenlines, flushbottom]{revtex4-1}
\usepackage{xspace,amsmath,amsfonts,amsthm,amssymb,amsbsy,graphicx,color}

\newcommand{\ms}[1]{\mbox{\scriptsize #1}}

\begin{document}

\newtheorem{theo}{Theorem} \newtheorem{lemma}{Lemma}

\title{The role of quantum measurements in physical processes and protocols}

\author{Benjamin Cruikshank} 
\affiliation{U.S. Army Research Laboratory, Computational and Information Sciences Directorate, Adelphi, Maryland 20783, USA} 
\affiliation{Department of Physics, University of Massachusetts at Boston, Boston, MA 02125, USA}
\author{Kurt Jacobs} 
\affiliation{U.S. Army Research Laboratory, Computational and Information Sciences Directorate, Adelphi, Maryland 20783, USA} 
\affiliation{Department of Physics, University of Massachusetts at Boston, Boston, MA 02125, USA}
\affiliation{Hearne Institute for Theoretical Physics, Louisiana State University, Baton Rouge, LA 70803, USA} 

\begin{abstract}   
In this mainly pedagogical article, we discuss under what circumstances measurements play a special role in quantum processes. In particular we discuss the following facts which appear to be a common area of confusion: i) from a fundamental point of view measurements play no special role whatsoever: all dynamics that can be generated by measurements can be generated by unitary processes (for which post-selection is no exception); ii) from a purely \textit{physical} point of view, measurements are not ``outside'' of quantum mechanics; iii) the only difference between the abilities of measurement-based protocols and unitary circuits for quantum computing comes from practical (technology dependent) constraints. We emphasize the importance of distinguishing between differences that are i) fundamental but without physical import; ii) fundamental and possess physical import; iii) are not fundamental but have practical import. We also emphasize the importance of separating theoretical and experimental elements of measurement, primarily \textit{projection} and \textit{amplification}, which are physically very different. Note that since we are concerned with facts regarding physical processes, this article has little if anything to do with interpretations of quantum mechanics. 
\end{abstract} 


\maketitle 

\section{Introduction}
\vspace{-3mm}

Quantum measurement theory, by which we mean the projection postulate and the formalism that follows from it, has an odd history; for decades physicists tried to avoid it, motivated by the curious philosophical issues that it raises~\cite{Schrodinger35, Schrodinger35t, Einstein35, Einstein36, Landau65, Gillespie86}. As experiments advanced, however, the action of measurements on individual systems in real-time became impossible to ignore, and quantum measurement theory was fully embraced in the quantum optics, quantum information, and related communities~\cite{mikeandike, WM10, Jacobs14}. In taking measurement theory on board, it appears that practitioners generally adopted the point of view that quantum measurements introduce into quantum mechanics some element that is otherwise absent. 
Recent conversations with a number of physicists lead us to believe that there was, and still is, widespread confusion as to what this element is. Indeed it is not so surprising that such a confusion might exist, given the ways that measurements tend to be referred to in the literature~\footnote{The following passages taken from papers in the areas of fault-tolerant quantum computation and the quantum-to-classical transition illustrate the way that measurements are often referred to: ``... we make an effort to show that fault tolerance can be achieved without using classical operations or measurements. We do this because of two reasons. One is purely theoretical: One would like to know that measurements and classical operations are not essential and that the quantum model is complete in the sense that it can be made fault-tolerant within itself.''; ``... we need to add a new component, and that component will be measurement.''; ``... we find that continuous measurement can effectively obtain classical mechanics from quantum mechanics.''.}.  
  

In fact, as far as dynamical processes are concerned --- that is, anything that can be achieved by any physical process or protocol --- measurements introduce nothing new, and are thus not ``outside'' of quantum mechanics. Any protocol that employs explicit measurements can be rewritten entirely in terms of unitary processes, and is thus contained within the theory of quantum mechanics. Similarly any experiment that uses measurements to generate some physical process can be reproduced by an experiment that uses only unitary processes (with the sole exception of a single measurement performed at the end of the process, a measurement that does not affect the dynamical description). If a physical measurement-based process is designed to produce a single outcome, then even the measurement at the end is unnecessary. These statements follow from the fact that the so-called ``measurement problem'' of quantum theory has no impact on the physical predictions of the theory. This does not mean that the measurement-based and unitary versions of an experiment will be equally easy to implement --- that depends on the available technology. 

Nevertheless, practical considerations motivate problems that involve constraints. These constraints may distinguish in some way between measurements and unitary processes so as to render them inequivalent as far as the problem is concerned. But the only way to distinguish between the two, from a physical point of view, is to expand the definition of measurement so that a ``measurement'' does more than merely select out one of a set of mutually orthogonal subspaces. A prime example of a problem that involves practically-motivated constraints is fault-tolerant quantum computation (FTQC), in which quantum gates, the source of unitary operations, come with an error probability. In this case the definition of measurement is expanded to include \textit{amplification}. Amplification is the act of taking a set of states that have only small differences in energy and transforming them into a set of states in which the energy differences are much larger. The importance of amplification is that it allows the information encoded in the set of states to be stored and processed on a classical computer. The larger energy scale is what allows the classical computer to store and process information error-free. Since systems with a larger energy scale are harder to protect from decoherence, amplification is associated with a loss of coherence in a given basis, and this is of course entirely compatible with making measurements. (The fact that classical computers are always decohered in a given basis is the reason that one equates the conversion of information from a mesoscopic quantum device to a classical computer with a measurement process.)  For quantum computation, since mesoscopic quantum gates must maintain coherence, it is natural to restrict these gates so that they cannot include amplification. Since measurement is now the sole provider of amplification, and amplification is a physical process (projection is not), measurement now brings a unique physical element to the problem and thus has a special role. This special role is to allow classical information (information stored in a given basis) to be processed error-free.  

In fact, as we discuss below, amplification is not required for error-free classical processing. This can be achieved with mesoscopic (and thus error-prone) gates by using them to construct fault-tolerant circuits. In view of this, unitary circuits can provide all the properties of measurement-based protocols that we have considered so far:  measurement, processing of the classical measurement results, and feedback consisting of operations applied to the computational qubits. Given that there may be technological advantages in implementing FTQC without measurements~\cite{DiVincenzo07, Paz10}, it is an interesting question as to precisely what effective measurement errors and feedback errors unitary circuits can achieve when used to replace measurements, for a given gate error. The above fact also begs the question as to why all high-threshold FTQC schemes developed to-date involve measurements. We devote some space in what follows to these two questions.  

There is one further special property with which measurements (actually amplification) are commonly imbued in analyses of FTQC schemes. This special property is the assumption that macroscopic classical gates are much faster than quantum gates. That is, a mesoscopic quantum gate that performs a given classical logic operation (AND, OR, NOT, \textit{etc}) is assumed to take significantly longer than the equivalent macroscopic gate. This confers on measurements the ability to provide much faster classical processing than possible with unitary circuits. Although this assumption may be true of present-day devices, it is not clear that it is justified in the longer term. While the energy scale of macroscopic systems is (by definition) larger than that of mesoscopic quantum systems, and this difference might lead to a fundamental gap between the speed of mesoscopic and macroscopic gates, it is not clear that it does so. Certainly the fundamental timescale of mesoscopic superconducting qubits ---  the separation of their energy levels --- is in the GHz range, and thus similar to the clock speed of classical electronics~\cite{Vijay12, Lecocq15, Kelly15, Ofek16, Huang14}. 

In the following we explain in detail, and thus elucidate and justify, the various statements and claims made in the above discussion. We explain how measurement-based processes are rewritten as unitary processes, especially those involving post-selection for which this translation is less obvious. Two areas in which the relationship between measurement-based (those that employ amplification) and purely unitary (purely mesoscopic) protocols are most relevant are those of FTQC and quantum feedback control. For the former we discuss in some detail how purely mesoscopic schemes can be designed to reproduce the behavior of amplification-based schemes. For the latter  we consider a range of open questions that are somewhat different from those posed by FTQC. We also elucidate  the difference between the ``measurement problem'' and the question of the physical properties of measurements. 

%
%
%
%

\vspace{-2mm}
\section{Physical processes} 
\vspace{-2mm}

A specification of the way in which the state of a physical system changes with time is called a \textit{dynamical process}. We will refer to a dynamical process as a \textit{physical process}: the state of a system is a complete description of its physical properties, and so a physical process is something that describes how the physical properties of a system change with time. Quantum Mechanics is a probabilistic theory, so our physical processes must be able to describe random changes in the state of a system. We therefore define a physical process as follows: it is something that, given an initial state at time $t$, specifies all the possible states at a time $t' > t$, as well as the probability with which each will occur. Central to our discussion here is the fact that a \textit{physical process is completely defined by the final states and the probabilities of these states.} The question of \textit{which} of the final outcomes actually occurs is irrelevant for the purposes of specifying the process. Specifying which of the outcomes occurs is equivalent to simulating a single ``run'' of the process, but since one does not know which realization of the process will occur before the run, and all the possible realizations are already specified by the process, the mere act of simulating a run does not provide any further information about the physical behavior. 

In quantum mechanics a physical process, which is generated by the fundamental forces (the laws of physics), is specified by a unitary transformation, along with a partitioning of Hilbert space into a set of mutually orthogonal subspaces. The subspaces define the mutually exclusive outcomes, and the probabilities of these outcomes are given by the Born rule: they are the square moduli of the projections of the final state of the system onto each of the subspaces. The unitary transformation and the set of subspaces completely determines the physical process. In fact, to be more precise, the process is already fully defined by the unitary transformation alone (and thus by the laws of physics): this unitary transformation tells us, for every way in which we may choose to break up Hilbert space into a set of mutually exclusive outcomes, the probabilities of these outcomes. 

To look ahead a little, quantum measurement theory is the machinery --- consisting essentially of sets of projectors --- that actually performs the projections onto the mutually exclusive subspaces, and thus picks realizations of the physical process. The ``use of measurements'' in any protocol or process means merely that one is explicitly introducing projection operators so as to describe the process in terms of an individual realization (the state projected onto one of the subspaces). The operation of the protocol is then described as an explicit stochastic process with classical probabilities. One can alternatively remove the projection operators from the analysis, so that the evolution is described in the full Hilbert space that contains all the possibilities simultaneously. In that case no explicit classical probabilities need be involved, except to record the probabilities of the final outcomes of the protocol (if needed). But in both cases the physical process is the same: the only role of measurement is to select out a subspace and thus pick a realization. For this reason, from a fundamental point of view, all protocols that involve measurements are automatically also protocols that are purely unitary. The two kinds of protocols are inseparable. 

\vspace{-2mm} 
\subsection*{Where is the ``measurement problem''?}
\vspace{-2mm} 

The ``measurement problem'' arises from the fact that the unitary evolution of quantum mechanics can, in general, place systems in superpositions of any chosen basis of orthogonal states. Since mutually orthogonal states can be distinguished by an observer, upon doing so an observer must find the universe to be in one of these states, and not a superposition. However, the evolution of quantum mechanics does not provide any dynamical process (any evolution via a fundamental force) that ``collapses'' a superposition of a set of states to one of these states when a measurement is made. As a result measurement is described merely as a process in which the possible states of an observer become  correlated with those of the system being measured. The Born rule, which connects quantum states to our observed reality, says that the observer will find the universe to be in one of a set of possible orthogonal states and determines the probability for each one. Because the observer's possible states are correlated with a set of states of the system, if we select one of these states as reality we necessarily select one state for the system, and the observer will find the system to be in that state. 

Thus quantum theory determines the possible states that will be observed, and the respective probabilities for these states, but provides no mechanism where by one of the states actually becomes reality, since the joint system of the observer, the system, and the rest of the universe continues to exist in a superposition of all the possibilities. This description of the universe is at variance with our everyday experience: quantum mechanics asserts that the universe is \textit{actually} in a superposition of a certain set of possible orthogonal states, while we \textit{actually} observe the universe to be in just one of these states. As pointed out above, however, this conundrum (usually referred to as the ``quantum measurement problem'') is philosophical, not physical, since it is separate from the question of the predictions of the theory as to the dynamics of physical processes: a physical process is defined only by the set of possible states and their respective probabilities. We are therefore not concerned in this article with the quantum measurement problem. 

\vspace{-2mm} 
\section{Unitary implementations of measurement-based feedback}
\vspace{-2mm} 
\label{uimp}

To gain insight into how measurement-based protocols map into unitary protocols it is useful to consider a simple measurement-based feedback procedure. A measurement on a quantum system $\mathcal{S}$, whose state is given by the density matrix $\rho$, is described by a set of ``measurement operators'' $\{A_n\}$, in which the final state of the system for outcome $n$ is given by 
\begin{equation}
   \tilde{\rho}_n = \frac{A_n \rho A_n^\dagger}{P_n} , 
\end{equation}
where 
\begin{equation}
   P_n = \mbox{Tr}[A_n^\dagger A_n \rho ]  
\end{equation}
is the probability that this outcome occurs. Once an observer has obtained the measurement result (the value of $n$), then he or she can apply a unitary transformation to the system that depends on this result. Denoting each of these unitary operations by $U_n$, the complete dynamics resulting from the measurement and the ``feedback'' operation is 
\begin{equation}
   \tilde{\rho}_n' = \frac{U_n A_n \rho A_n^\dagger U_n}{P_n} . 
\end{equation}

We can implement the above measurement and feedback operation unitarily using a second ``auxiliary'' quantum system, $\mathcal{A}$, with dimension $n$. Denoting a basis for $\mathcal{A}$ by $\{ |n\rangle_{\ms{a}} \}$, we first prepare $\mathcal{A}$ in the state $|0\rangle_{\ms{a}}$, and then apply a joint unitary $V$ to $\mathcal{S}$ and $\mathcal{A}$. By choosing $V$ appropriately we can arrange that the final joint state of the two systems is 
\begin{equation}
   \sigma =  V (\rho \otimes |0\rangle_{\ms{a}} \langle 0 |_{\ms{a}} )= \sum_{n,m} A_n \rho A_m^\dagger \otimes |n\rangle_{\ms{a}} \langle m |_{\ms{a}} . 
\end{equation} 
The relationship between the set of operators $\{ A_n\}$ and the required unitary $V$ is quite simple (see, e.g.~\cite{Jacobs14}). The state $\sigma$ is one for which the system $\mathcal{S}$ is in the state $\tilde{\rho}_n = A_n \rho A_n^\dagger/P_n$ when $\mathcal{A}$ is in the state $|n\rangle$ . To see this, first we separate out the diagonal terms in the sum: 
 \begin{equation}
   \sigma = \sum_{n} A_n \rho A_n^\dagger \otimes |n\rangle_{\ms{a}} \langle n |_{\ms{a}} + \sum_{k\not= m} A_k \rho A_m^\dagger \otimes |k\rangle_{\ms{a}} \langle m |_{\ms{a}}. 
\end{equation} 
If we want to examine the state of $\mathcal{S}$ when $\mathcal{A}$ is in state $|n\rangle_{\ms{a}}$, then we need to examine the joint density matrix restricted to the subspace in which $\mathcal{A}$ is in state $|n\rangle_{\ms{a}}$. This is given by the term in the expression for $\sigma$ above that contains only the outer product $|n\rangle_{\ms{a}} \langle n|_{\ms{a}}$. The state of the system in this subspace is the coefficient of $|n\rangle_{\ms{a}} \langle n|_{\ms{a}}$, and is thus $A_n \rho A_n^\dagger$. This state is not normalized, however. To normalize it we divide it by its trace with is $\mbox{Tr}[A_n^\dagger A_n \rho] = P_n$. The result is the state $\tilde{\rho}_n$ defined above. 

To determine the probability that $\mathcal{A}$ is in state $|n\rangle_{\ms{a}}$ we use the Born rule, and thus calculate the trace of the state obtained by projecting $\sigma$ into the subspace defined by $|n\rangle_{\ms{a}}$. Since this projected state is just $A_n \rho A_n^\dagger$, this trace is $\mbox{Tr}[A_n^\dagger A_n \rho] = P_n$. 

To summarize the above analysis, if we define the outcomes of our physical process by partitioning the Hilbert space into the subspaces given by the states $\{ |n\rangle_{\ms{a}}\}$ of $\mathcal{A}$, then the state $\sigma$ precisely encodes the physical state corresponding to having the outcomes $\tilde{\rho}_n$ occurring with probability $P_n$, and thus to the physical process given by the measurement with operators $\{ A_n \}$. 

To apply to $\mathcal{S}$ the feedback unitaries $U_n$ all we have to do is to apply a joint unitary to the two systems that applies the unitary $U_n$ to the system when $\mathcal{A}$ is in state $|n\rangle_{\ms{a}}$. This unitary is 
\begin{equation}
   W =  \sum_{n} U_n \otimes |n\rangle_{\ms{a}} \langle n |_{\ms{a}} . 
\end{equation} 
Applying $W$ to $\sigma$ the joint state of the two systems becomes  
\begin{equation}
   \sigma' =  \sum_{n} U_n A_n \rho A_m^\dagger U_m^\dagger \otimes |n\rangle_{\ms{a}} \langle m |_{\ms{a}} ,  
\end{equation} 
which, as we have seen above, describes the physical situation in which the system is in one of the states $ \tilde{\rho}_n'$ with the corresponding probabilities $P_n$. From a physical point of view the unitary process is thus equivalent to the measurement-based feedback process above, in that an observer who measures in the basis $|n\rangle_{\ms{a}}$ will see the same outcomes with the same probabilities. 


\vspace{-2mm} 
\subsection*{Physical processes with a single outcome} 
\vspace{-2mm} 

Often the purpose of a protocol is to produce a single final state rather than a number of possibilities. In such protocols, especially those that must deal with random external influences, there may be a large number of intermediate possibilities that must somehow be reduced to a single final outcome. If we wish to reduce all outcomes of all subsystems taking part in our process to a single outcome, then this is only possible if we reverse the evolution of the external systems that have generated the multiple possibilities. But so long as we have enough auxiliary systems taking part in our process then we can always place the system of interest in a single final state, no matter the number of intermediate possibilities.  This can be done, for example, by using an auxiliary system to implement a feedback process similar to the one described above. If system $\mathcal{S}$ is in a mixture of some set of orthonormal states $|n\rangle_{\ms{s}}$, each having probability $p_n$, then $\mathcal{S}$ can be transformed into a single pure state $|\psi\rangle_{\ms{s}}$ in the following way. First we initialize the auxiliary system in the state $ |0\rangle_{\ms{a}}$. Then we define a unitary operator 
\begin{align}
  U = \sum_{n} |n\rangle_{\ms{s}} \langle n |_{\ms{s}} \otimes |n\rangle_{\ms{a}} \langle 0 |_{\ms{a}} + \tilde{U}
\end{align} 
designed to correlate system $\mathcal{S}$ with the auxiliary. Here the operator $\tilde{U}$ is chosen merely to ensure that $U$ is unitary. For example we could set 
\begin{align}
  \tilde{U} = & \; |0\rangle_{\ms{s}} \langle 0 |_{\ms{s}} \otimes \sum_{n\not=0}  |n\rangle_{\ms{a}} \langle n |_{\ms{a}}  \nonumber \\ 
  & \; +  \sum_{n\not=0}  |n\rangle_{\ms{s}} \langle n |_{\ms{s}} \otimes \biggl(  |0\rangle_{\ms{a}} \langle n |_{\ms{a}} +  \sum_{k\not=0,n}  |k\rangle_{\ms{a}} \langle k |_{\ms{a}}  \biggr) . 
\end{align}
Now we apply $U$ to the two systems. Since the auxiliary is initially in the state $ |0\rangle_{\ms{a}}$ the operator $\tilde{U}$ has no effect and the result is the correlated state  
\begin{align}
  \sigma_U   =  \sum_{n} p_n |n\rangle_{\ms{s}} \langle n |_{\ms{s}} \otimes |n\rangle_{\ms{a}} \langle n |_{\ms{a}} . 
\end{align}
We now define the unitary operator  
\begin{align}
  V   =  \sum_{n} |\psi\rangle_{\ms{s}} \langle n |_{\ms{s}} \otimes |n\rangle_{\ms{a}} \langle n |_{\ms{a}}  + \tilde{V}, 
\end{align}
in which, once again, $\tilde{V}$ is chosen merely to ensure that $V$ is unitary.  Applying $V$ to the two systems implements the ``feedback'' process and produces the final state  
\begin{align}
  \sigma_{\ms{fin}}   =   |\psi\rangle_{\ms{s}} \langle \psi |_{\ms{s}} \otimes \Bigl[ \sum_{n} p_n |n\rangle_{\ms{a}} \langle n |_{\ms{a}} \Bigr] . 
\end{align}
Note that this feedback process actually transfers the initial state of the system into the auxiliary. Since the entire process is unitary (and thus reversible), it must also transfer the initial state of the auxiliary into the system (up to a local unitary transformation), which it does. We could, of course, combine the two unitary operations together to obtain the ``swap'' unitary $S = VU$. Here we implemented the ``swap'' in two parts to show how it is achieved by using a correlating process (essentially a measurement) followed by a ``feedback'' process. Since all basic physical processes are logically reversible, to reset a system to a predefined pure state one must always swap the entropy in the system to another system. This is the origin of Landauer's erasure principle, in which erasing the information in a bit (reseting a bit) requires raising the entropy of the environment~\cite{MyEprint, Jacobs14}.

\vspace{-2mm} 
\section{Translating measurement-based protocols into unitary protocols} 
\vspace{-2mm} 

In the analysis above we showed how any physical measurement-based feedback process can be implemented with a purely unitary process. One can clearly construct a unitary protocol that reproduces \textit{any} protocol involving measurements merely by using the above equivalence to replace all occurrences of a measurement with a unitary process involving an auxiliary quantum system. Nevertheless, it appears to be a natural tendency to assume that the unitary processes that result from replacing measurements with auxiliary systems would be much more complex and cumbersome to implement than the equivalent measurement-based protocols because of the need to ``carry around'' all the Hilbert space and use all those auxiliary systems. However a simple line of reasoning shows that this is not the case.  

Consider a unitary protocol $\mathcal{U}$ that performs the same physical process as a measurement-based protocol $\mathcal{M}$. The parts of $\mathcal{U}$ that implement each of the measurement and feedback subprocesses, in the manner described in Section~\ref{uimp} above, do not require coherence between the states of the basis used for the auxiliary. We can thus allow the all the auxiliary systems to decohere in the appropriate bases. This decoherence does not change the complexity of the implementation. It does however, render the unitary implementation identical to the measurement-based implementation. The auxiliary quantum systems can now be thought of as classical systems performing the feedback operations. In fact, the unitary protocol is now an exact description of the measurement-based protocol, and as such the unitary protocol is no more complex that the measurement-based one. 

In a measurement-based protocol the macroscopic classical systems that perform the measurement and feedback operations are continually being reused. The information contained in the classical bits is thus repeatedly erased, which ultimately involves dumping this entropy to an environment. In exactly the same way, the qubits in a unitary protocol can be reused by swapping their states into a cold reservoir to reset them to zero. Since parts of a unitary protocol may be decohered and/or erased by interactions with a bath, doing so is likely to be the simplest way to perform the protocol. The resulting evolution is still unitary, but if one traces out the bath(s) involved the resulting description of the protocol is no-longer unitary. To avoid confusion that might arise from referring to such protocols as unitary, we will instead use the term ``measurement free''. This emphasizes the key property, which is that the protocol does not use measurements, without concerning ourselves with whether baths are used to allow reset operations or induce decoherence. 

Even given that unitary and measurement-based protocols can have the same structure, it is not immediately clear from the above analysis how post-selection protocols are implemented unitarily. We discuss the implementation of these protocols in next two sections.   
 
\vspace{-2mm} 
\subsection{``Static'' vs. ``Dynamic'' unitary circuits} 
\vspace{-2mm} 
 
Post-selection is the term used for a process in which a collection of systems is prepared in some joint state, some of the systems are measured, and depending on the measurement result the rest of the systems are either used as part of an ongoing process, or discarded. If the rest of the systems are discarded, then the process is repeated until the measurement returns a result that allows the systems to be used. In this way an iterative process with a variable number of iterations is used to prepare quantum states that are required for some process, such as a fault-tolerant quantum computation. 

Unitary implementations of quantum computing are assumed to use a circuit (meaning a specified sequence of quantum gates applied to a number of qubits) that is fixed at the start of the computation and then implemented. Post-selection, however, involves running a unitary circuit to prepare the state of a collection of qubits, and then deciding whether or not to run the circuit again based on the result of a measurement. These two kinds of implementation seem somewhat different, so let us refer to an implementation that involves a fixed gate sequence as a ``static'' circuit, and one in which the gate sequence is changed as the process proceeds based on (random) measurement outcomes as a ``dynamic'' circuit. 

A first sight it may seem that the amplification of measurement provides a unique ability to modify the gates performed on qubits because these gates are implemented with macroscopic controls. These controls apply macroscopic forces to the qubits to implement the gates in the fastest way. We cannot merely replace the measurement and feedback processes that choose the gates with mesoscopic unitary processes because the mesoscopic elements cannot control the macroscopic elements that implement the gates.  But let us suppose for a moment that we do wish to use a mesoscopic qubit $Q$ to control which gates are applied to and between other qubits. Since quantum gates are typically two-qubit gates, having the action of these gates controlled by another qubit merely means replacing the two-qubit gates with three-qubit gates. We then arrange things so that all the gates in the circuit to be controlled are controlled by the same qubit, namely Q. But we also know that two qubit gates are universal for unitary computation, so all the three qubit gates can be replaced by two-qubit gates. There is therefore no problem at all with having Q control the action of gates between other qubits. The complexity of the circuit has been increased somewhat (there are now more two-qubit gates than there were before), but the additional complexity is presumably no more than the complexity of the classical controller that would otherwise have to act based on the results of a measurement. 

Now note that to implement post-selection by having one or more qubits, Q, control whether or not a given circuit, C, is applied to some other qubits, we do need the ability to run circuit C repeatedly. We can, of course, do this merely by continually repeating the fixed sequence of gates that constitute C. While this gate sequence is fixed, the action of the gates, and thus the circuit C, is different each time because this action depends on the value of the control qubit(s) Q. The circuit C is part of a larger circuit L that is also run each time C is run. The circuit L includes the gates that correlate the control qubit(s) Q with a subset of the other qubits (so that Q effectively stores the result of a measurement on these qubits), and thus updates the value of Q each time it is run. We now have a perfectly deterministic circuit, L, that is able to implement a repeat-until-success post-selection protocol. It does so by virtue of the fact that its action on each repetition depends on information stored on previous repetitions. In fact, by introducing a process of repeating a fixed circuit an indefinite number of times to perform a computation we are merely employing the procedure already used by classical computers. It is precisely this method of operation that allows classical computers to perform universal computation (and thus post-selection) with fixed circuits. We see that the distinction between ``static'' and ``dynamic'' unitary circuits is merely that the former consist of a fixed sequence of gates applied once, whereas the latter consist of a fixed sequence of gates applied repeatedly until the computation is complete.

\vspace{-3mm} 
\subsection{Post-selection with a fixed-time circuit} 
\vspace{-1mm}

In the previous discussion we saw that unitary circuits can be used to perform post-selection processes in the same ``dynamic'' repeat-until-success manner employed by measurement-based protocols. However, we feel it is worth noting that one can also perform post-selection processes with ``static circuits'', meaning a gate sequence that is performed only once and that does not involve the repetition of state preparations. This is achieved by having all the preparations done simultaneously at the start (in parallel), which requires effectively fixing the number of repetitions that the post-selection process is allowed to use. Fixing the number of repetitions does not make a significant difference to the output of the protocol for two reasons. The first is that the number of repetitions that a dynamic procedure can use to prepare a state for use in a computation is in any case limited by practical considerations. The second is that the probability that the procedure does not produce a correct state for the computation reduces exponentially with the number of repetitions, so in practice only a small number is required. 

Let us consider an example in which the post-selection protocol involves i) preparing two identical states of $m$ qubits each, ii) applying gates that interact the combined set of $2m$ qubits, iii) performing a measurement on one of the $m$-qubit subsets, and iv) keeping or discarding the other subset depending on the result of the measurement. To perform this protocol with a fixed circuit we apply $N$ copies of the circuit that performs steps i) and ii) simultaneously on $N$ respective sets of $2m$ qubits. We then mimic step iii (the measurement) by applying (again simultaneously) to each of the $N$ sets a circuit that correlates one of their two $m$-qubit subsets a single control qubit (each of the $N$ sets has its own control qubit, which stores the result ``keep'' or ``discard''). We now construct a register of $m$ new qubits, R, that will contain the output of our protocol. We then go to each of the $N$ sets in turn, and perform a swap operation that swaps its unmeasured subset of $m$ qubits into R, conditional on the state of the control qubit for that set. In this way, each of the $m$ subsets is loaded into the output register in turn, but only if it contains a state that has passed the ``measurement test''. The larger the value of $N$ the less likely it is that the register does not contain the desired output state, with the likelihood dropping exponentially in $N$. The action of this circuit is the same as that of the measurement-based repeat-until-success protocol with the maximum number of repetitions fixed at $N$ (except for the time that the output states have to spend sitting in the output register, which is linear in $N$). 

\vspace{-2mm}
\subsection{Unitary error-free classical processing}
\vspace{-1mm}

To fully replace measurements with unitary circuits we must have an explicit construction by which the unitary circuits can perform perfect (that is, fault-tolerant) classical computation. This could be achieved by using the concatenation methods of FTQC, but these have a component-wise error threshold of around $1\times 10^{-3}$~\cite{Antonio16x}. Further, the information that we wish to process, which is stored in the computational basis states of mesoscopic qubits, must be encoded first so that fault-tolerant processing can be performed on it. We must therefore have an explicit circuit for performing this encoding, and consider the probability that this circuit encodes the state incorrectly. This error probability for encoding is precisely equivalent to the measurement error in a measurement-based protocol. We recently presented an explicit scheme for encoding classical information and processing it reliably using unitary gates~\cite{Cruikshank16x}. This method builds upon von Neumann's ``multiplexing'' method~\cite{vonNeumann56}, appears to be quite feasible to implement (unlike previous multiplexing protocols), and achieves an error threshold close to von Neumann's ideal value of $\epsilon=1/6$ for a "majority organ" (see below) in an error-correcting network. We now briefly summarize the performance of this method. 

\begin{figure}[t]
\leavevmode\includegraphics[width=1\hsize]{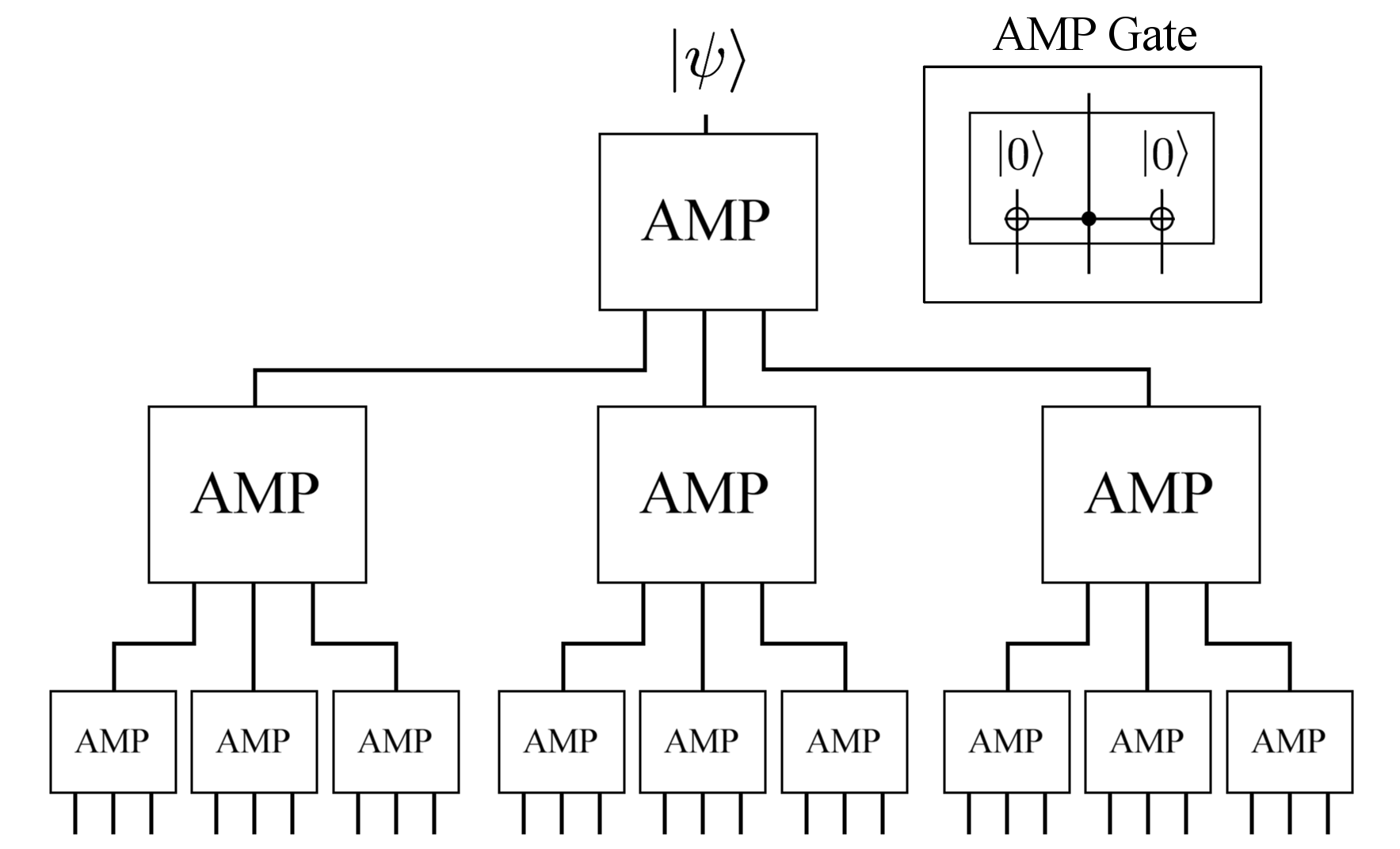}
\caption{A circuit that encode a single classical bit (which may be stored in a qubit) into a repetition code of $3^n$ qubits, in which the encoding error is less than $0.51$ of the error probability of the AMP gate. We also show the AMP gate in terms of controlled-not gates. Of course, the bits in the resulting code contain mutual correlations. These correlations can be effectively eliminated (for a subset of $3^{n-1}$ bits) by using majority counting gates to realize a stable error-correction circuit. The result is a repetition code with effectively $3^{n-1}$ bits.} 
\label{fig1} 
\end{figure}

Our encoding and processing circuits are based on the simplest classical coding scheme, namely repetition coding, in which the state of the qubit is simply stored in some number of qubits. The state can be faithfully recovered so long as more than half the qubits retain the correct state. The circuit shown in Fig.~\ref{fig1} achieves the encoding process by copying the state of a single qubit to $3^n$ qubits by using a ``cascade'' of ``AMP'' gates, in which each AMP copies the classical information in a single qubit to two others (using, e.g., two CNOT's). The probability that there is an error in the resulting encoding is $\approx 0.51 p$, in which $p$ is the probability that there is a (quantum) error in at least one of the outputs of the AMP gate~\cite{Cruikshank16x}. Thus the equivalent measurement error incurred by this encoding circuit is $\approx 0.51 p$, and thus less than that of an individual AMP gate. 

To process the information in the repetition code fault-tolerantly we must have an error correction procedure (that is, a unitary circuit) which is applied periodically to the $3^n$ bits in the code. To achieve this error correction von Neumann used a ``majority organ'' that takes three input bits and, so long as it works correctly, outputs three bits with the value shared by the majority of the inputs~\cite{vonNeumann56}. We construct this irreversible three-bit gate with two three-bit unitary gates- a reversible majority counting gate, and an AMP gate to copy the correct output to two ancillary bits. Applying the majority organ to triples of the $3^n$ code bits certainly corrects errors, but it also produces correlations between the bits belonging to each triple, degrading the code. Von Neumann got around this problem by randomly regrouping the bits into new triples each time one desires a correction operation. In this way, so long as there are enough bits in the code, the bits input to each majority gate remain sufficiently independent to achieve a stable steady-state. Nevertheless the procedure demands highly complex wiring with the result that its scalability is unclear~\cite{Pippenger90, Han11} (see also~\cite{Nikolic02, Roy05, Boykin05, Beiu07, Bhaduri07}). It turns out that it is possible to use majority gates in a manner that is much simpler and has a performance which is quite similar. By i) arranging the $3^n$ code qubits in an $n$-dimensional hypercube, ii) defining a single correction operation as that of sequentially applying majority gates along each of the $n$ dimensions of the hypercube in turn, and iii) defining the code ``bits'' as the $3^{n-1}$ groups containing three mutually correlated bits, one obtains a high-threshold fault-tolerant correction circuit that is highly efficient in that it enables a compact (wiring) configuration~\cite{Cruikshank16x}. As an example, an 81-(qu)bit code (n=4) turns a three-bit unitary gate error of $ p = 0.4\%$ into a logical error probability of $1.5\times 10^{-11}$.  

Finally we must have a way to process the information stored in the repetition code. This is not difficult, since a logical operation on a logical (coded) bit is obtained merely by applying that operation to each of the coding (qu)bits. In fact, the AMP and majority gates are universal for classical computation. Since all inputs of a three-input gate must be correct in order to produce a correct logical result, the threshold is slightly smaller for universal computation and was calculated to be $p = 5.5\%$~\cite{Cruikshank16x} or $\epsilon\approx 13\%$.

\subsection{Summary: measurement-based discrete feedback vs. a unitary implementation} 

To reproduce measurement-based feedback operations using unitary circuits requires encoding the information, processing it, and then using the result to apply an operation to one or more qubits (the ``feedback''). We discussed the first two above. To apply the feedback operation to some ``target'' qubit we can simply use any one of the code qubits, since the error-correction circuit maintains each on of them as a copy of the encoded logical bit, with some (steady-state) probability of error. Using the error-correction circuits described above, this probability is $q \approx 0.51 p$. If the feedback operation is then performed by applying a two-qubit gate to the chosen code qubit and the target qubit, the total probability of error is that of the two-qubit gate (which we take to be $p$) plus that of the coding qubit, and is thus $\approx 1.51 p$.  

We can now summarize the equivalent measurement error and feedback error achievable with unitary circuits using majority gates with error probability of $p$. The equivalent measurement error (as discussed above) is no more than $0.51 p$, the processing has an arbitrarily low error, and there is an error on the feedback operation of $0.51 p$ in  addition to the basic gate error (which is that incurred by the equivalent classical feedback operation). To put this succinctly, unitary circuits can implement any measurement and feedback operation, and do so with an error probability very similar to the unitary gate error probability $p$.  

\subsection{Fault-tolerant quantum computation} 

In their analysis of FTQC, Aharanov and Ben-Or used an explicitly measurement-free implementation~\cite{Aharonov98, Aharonov08}. Their purpose was to demonstrate that scalable FTQC was possible, and not to provide a scheme with a high threshold, and their scheme had a threshold of $p \approx 10^{-12}$. Since then quite a few fault-tolerant schemes have been devised with the purpose of obtaining higher thresholds~\cite{Shor96, Steane99, Knill05, Aliferis06b, Raussendorf07a, Svore07, Fowler09, Fujii10, Paetznick13, Stephens14}, with the most recent claiming a threshold of $p \approx 0.01$. All of these latter FTQC schemes employ measurements, thus begging the question as to whether measurements are required for high threshold FTQC. As we have explained above, there can be no difference between measurement-based and unitary schemes that originates from quantum theory itself; any difference can only be due to technological factors. 

Now consider the technological difference that arises between measurement-based and unitary FTQC schemes when measurements come with the ability to perform amplification. We have discussed above the fact that any amplification-based process that forms part of an FTQC protocol can be replaced by mesoscopic unitary circuits, and this can be done in such a way that the resulting error probabilities are very similar to those of the basic unitary quantum gates. Therefore, as far as error probabilities go, one expects the ultimate limits to the thresholds for unitary and measurement-based protocols to be similar. However, the error rate is not the only property of the feedback processes that is important for FTQC thresholds. The time required by these processes is also important. Any time that the computational qubits spend waiting for classical information to be processed incurs the accumulation of errors and thus affects the threshold. The relative thresholds of measurement-based and unitary FTQC protocols will thus depend, among other factors, on the relative time required by amplification as compared with encoding, something that again depends on technology. 

\section{Measurements vs. unitary quantum control: open questions} 

We have discussed the relationship between measurements and unitary processes from a fundamental point of view, and from a more practical (technological) point of view in the context of quantum computation. There is another area in which the practical relationship between measurements and unitary processes is especially salient, and that is \textit{feedback control}~\cite{Jacobs13, Jacobs14}. Quantum computation certainly uses feedback control, but when referring to feedback control as a subject in itself one usually has in mind applications of feedback for controlling the \textit{state} of a system (rather than to dynamical processes that enable the processing of information, as in FTQC). The difference is that the processing of information requires that the control protocol does not know the state of the system (it must operate correctly for any input state) whereas this is not the case if one wants merely to prepare a given state or realize a specific evolution. Because of this the problems typically studied under the term ``quantum feedback control'' tend to be much simpler as far as the protocols are concerned, but have other features that distinguish them from those in quantum computation.  

Feedback control is usually concerned with continuous-time processes, rather than the discrete gates of computation, and the relevant constraints are restrictions on the sizes of the available forces and the rates at which these forces can be changed.  The questions of interest tend to involve the fidelities that can be achieved under the relevant constraints in the presence of specified damping and/or noise sources. The problems are thus mainly of a dynamical nature rather than an information theoretic one. Nevertheless, since feedback protocols must usually transfer entropy from the system to the controller, environment, or other auxiliary systems, this area of research lies on the border between quantum dynamics, thermodynamics, and information theory~\cite{Strasberg13, Horowitz14, Brandner15, Horowitz15, Gong16}. 

As in quantum computation it is natural to define measurements as including amplification. Measurements thus bring with them error-free processing of the information obtained from the system. Further, depending on the technological constraints relevant to a particular problem, there may be some differences between the fidelity or speed of the control operations that can be applied to a quantum system by a macroscopic, as opposed to a mesoscopic, device~\cite{Balouchi16x}. 
The second difference between measurement-based controllers and mesoscopic quantum controllers is that measurement-based controllers are more restricted in the way in which they can interact with the system being controlled. As we have emphasized in this article, measurements do not provide any fundamentally new dynamics. In fact, the requirement that a controller  interact with a system via making measurements on it actually restricts the dynamics that the controller can employ. For example, the controller cannot use joint dynamical processes that generate entanglement between the system and the controller. 

We see that the question of the usefulness of measurements for control is somewhat different from that of their usefulness for FTQC.  While measurement (more precisely amplification) provides stronger control forces and error-free processing of the information obtained about the system, a controller that is a classical system, and whose interactions with the system must effectively involve measurements, cannot access the full range of dynamical interactions available to a controller that is a quantum system. When comparing the performance of measurement-based and coherent control one must also decide whether the constraint on the strength of the interaction between the system and the quantum controller should be the same as that for the probe system that makes the measurements for the measurement-based controller. Certainly setting these to be the same makes sense for present technologies, and the results we describe below are obtained under this assumption. 

To-date the question of the relative advantages of measurement-based and coherent feedback control is largely unexplored. Some results have been obtained for ideal controllers (that is, controllers that are not themselves subject to noise, and are able to make minimal-noise (quantum limited) measurements, etc.). It is known that ideal linear quantum controllers  outperform classical (measurement-based) linear controllers for controlling linear quantum systems~\cite{Hamerly12, Hamerly13, Jacobs14b}. It is known that when there is a bound on the maximum eigenvalues of the interaction between the system and the controller, a quantum controller can outperform a measurement-based controller~\cite{Jacobs14c}. It is also known that there are some tasks, especially for information processing, that quantum controllers can perform that measurement-based controllers cannot~\cite{Wiseman94d, Naoki14}. There has been virtually no research on regimes in which measurement-based control is superior to that by mesoscopic quantum systems; this may well be the case when mesoscopic controllers are subject to various kinds of noise. In fact there are many open questions regarding the operation and performance of coherent controllers and their relationship to measurement-based control. For classical controllers it is clear that the controller must obtain information about the system in order to effect control, and this information is easily quantified and its dynamics elucidated using established techniques. For quantum controllers the relationship between control achieved and information obtained is not at all clear; whether information is irrelevant to quantum controllers, or whether there are measures of quantum information that are relevant to quantum control is unknown. Exactly what mechanism(s) allow quantum controllers to outperform classical controllers has been scarcely elucidated, and whether there is any relationship with entanglement that may be generated between the controller and the system is unknown. In classical controllers the data obtained from measurements of the system must often be processed with complex circuits to effect good control, while those coherent control methods explored to date use virtually no processing. The value of processing to quantum controllers, or equivalently the value of increasing the complexity of quantum controllers has not been explored. The ultimate limits to control imposed by finite resources, such as interaction strength, are also largely unknown~\cite{Wang13, Balouchi16x}. 

\section{Conclusion}

We have elucidated in detail how measurements, defined as the action of projecting onto a subspace to obtain a measurement result, do not generate any physical effect that cannot be obtained via a unitary process. Because of this, no physical process or protocol can require the use of measurements for its implementation, unless by ``measurement'' one means more than the selection of an outcome along with any outcome-dependent actions. However, real measurements of mesoscopic systems in the laboratory always involve not only projection but also amplification. While projection does not  effect the physics of a process, amplification does since it is a physical process itself. 
We have also explained how amplification processes can be replaced by mesoscopic encoding processes, which shows that while amplification is physical it doesn't necessarily imbue measurement-based protocols with novel power. 

One way to summarize the main concepts we have discussed is to consider a statement that is fairly common in the literature, and has been referred to as ``the principle of deferred measurement''. This statement is ``measurements that are applied part-way through any physical process may always be removed and replaced by measurements at the end of the process''. If the reader has fully digested our discussion, we would expect that he or she might comment on this statement in the following way: ``If by measurement one means the application of projection operators in the manner of the standard quantum measurement formalism, then measurements at the end of the process are no more necessary than those in the middle; they are equivalent merely to defining the basis with which an observer would wish to correlate his or her measuring apparatus. The phrase `make a measurement at the end of the process' is somewhat misleading when referring only to applying a projector, since it does not involve any physical process and thus any action. If one is instead referring to measurement as the act of amplifying a signal (one of a set of quantum states) so as to correlate these states with a macroscopic measuring device, then from a practical point of view it is not necessarily the case that measurements can be removed from a process and placed at the end. Since amplification is a physical process, in general it will depend on the available technology as to whether the use of amplification as part of implementing a process is advantageous or not.'' 

We have also considered the question as to whether unitary protocols that maintain coherence have any unique power over that of measurement-based protocols. In the most general context the answer is yes, since quantum computing can perform tasks that classical computation cannot. Here we considered this question in the context of the dynamical control of quantum systems by other systems, often referred to as (measurement-based or coherent) feedback control. In this case it does appear that unitary protocols have some advantages over measurement-based protocols, at least in certain situations, but relatively little is yet known. As with the question of the relative advantages of measurement-base versus unitary protocols for implementing fault-tolerant quantum computation, it seems likely that in practice the answer will be dominated by technological rather than fundamental considerations. 

\vspace{2mm}
\section*{Acknowledgments} 
\vspace{-1mm}

KJ would like to thank Tanmoy Bhattacharya and Emil Mottola, of Los Alamos National Laboratory, for valuable discussions circa 2000 on the place of measurement in quantum physics. BC was supported by an appointment to the Student Research Participation Program at the U.S Army Research Laboratory, administered by the Oak Ridge Institute for Science and Education through an interagency agreement between the U.S Department of Energy and USARL.  
 

%

\end{document}